\documentclass[12pt]{article}
\usepackage{amsmath,amssymb,amstext,amsthm,exscale,latexsym}
\usepackage{array,graphicx,floatflt}
\newcommand {\al}   {\alpha}       \newcommand {\bt}  {\beta}
\newcommand {\g }   {\gamma}       \newcommand {\G }  {\Gamma}
\newcommand {\dl}   {\delta}       \newcommand {\e }  {\epsilon}
\newcommand {\z }   {\zeta}

\newcommand {\s }   {\sigma}      
         
\newcommand {\vf }  {\varphi}

\newcommand {\pl}   {\partial}     
       \renewcommand {\max}{{\sf\,max\,}}
\newcommand   {\const}{{\sf\,const}}     
     
\newcommand   {\sign}{{\sf\,sign\,}}     
\newcommand {\MM}  {{\mathbb M}}
\newcommand {\MR}  {{\mathbb R}}
\newcommand {\MS}  {{\mathbb S}}
\newcommand {\MU}  {{\mathbb U}}
\newcommand {\CC }  {{\cal C}}
\newtheorem{theorem}{Theorem}[section]
\begin{document}
\title     {Global solutions in gravity.\\
            Euclidean signature.}
\author    {M. O. Katanaev
            \thanks{E-mail: katanaev@mi.ras.ru}\\ \\
            \sl Steklov Mathematical Institute,\\
            \sl ul. Gubkina, 8, Moscow, 117966, Russia}
\maketitle
\begin{abstract}
We consider a wide class of two-dimensional metrics having one Killing vector.
The method is proposed for the construction of maximally extended surfaces with
the given Riemannian metric which is the analog of the conformal block method
for two-dimensional Lorentzian signature metrics. The Schwarzschild solution is
considered as an example.
\end{abstract}
\section{Introduction                                                  }
For physical interpretation of solutions in different gravity models, one has
not only to find the metric as a solution of the equations of motion but also
to analyze the behavior of extremals (geodesics) corresponding, in particular,
to trajectories of point particles. Therefore the construction of global
solutions is of uttermost significance. By global solution we mean a pair
$(\MM,g)$ where $\MM$ is a manifold and $g=\lbrace g_{\al\bt}\rbrace$ is a
metric given on $\MM$. The metric is to be found as a solution to some system of
the Euler--Lagrange equations and a manifold is supposed to be maximally
extended. The last requirement means that any extremal can be prolonged either
to infinite value of the canonical parameter in both directions or it ends up at
a singular point at a finite value of the canonical parameter where at least one 
of the geometric invariants is infinite or not defined. The well known example
is the Kruskal--Szekerez extension of the Schwarzschild solution
\cite{Kruska60,Szeker60}.

In general, this problem is very complicated because it requires an exact 
solution of the equations of motion as well as the analysis of extremals. The 
case of spherically symmetric solutions in general relativity was analyzed by 
Carter in \cite{Carter73}. The method of conformal blocks for constructions of 
global solutions for a wide class of two-dimensional metrics having one Killing 
vector was proposed in \cite{Katana00A}. This method was developed as  a result 
of construction and classification of all global solutions 
\cite{Katana90,Katana91,Katana93A} in two-dimensional gravity with torsion 
\cite{VolKat86,KatVol86,KatVol90}. The method of conformal blocks was also used 
for construction of global solutions in many two-dimensional dilaton gravity 
models \cite{KaKuLi96,KaKuLi97} and for complete classification of global vacuum 
solutions in general relativity with a cosmological constant assuming that the 
four-dimensional space-time is a warped product of two surfaces with a block 
diagonal metric \cite{KaKlKu99}. In the last case, all solutions were known 
locally. The analysis of global properties gave physical interpretation of many 
solutions which were earlier known only locally. Besides the black hole 
solutions, the vacuum Einstein equations have solutions describing cosmic 
strings, domain walls of curvature singularities, cosmic strings surrounded by 
domain walls, and other physically interesting solutions.

The method of conformal blocks is applicable for metrics of Lorentzian signature.
At the same time, the Euclidean formulation of the theory plays important role
in quantum field theory and statistical mechanics and allows in some cases to
avoid difficulties related to the indefinite signature of the metric. In the
present paper, the method of conformal blocks is generalized to a wide class of 
two-dimensional metrics of Euclidean signature having one Killing vector. We 
prove that there is the global solution for each conformal block with positive 
or negative definite metric. This method was applied earlier to two-dimensional 
gravity with torsion \cite{Katana97} and in the analysis of hyperbolically 
symmetric solutions in general relativity \cite{KaKlKu99}.
\section{Local form of the metric}
We start with the Lorentz case to explain the choice of the Riemannian metric
for the present paper. At first glance, the choice of Lorentzian metric may
seem artificial, but many exact solutions of general relativity depending
essentially on two coordinates can be written in this way. Besides, a general
solution of the equations of motion of two-dimensional gravity has exactly this
form in the conformal gauge \cite{Katana02}.

We consider a plane $\MR^2$ with Cartesian coordinates
$x^\al=\lbrace\tau,\s\rbrace$, $\al=1,2$. Two-dimensional metrics of constant
curvature as well as many solutions of general relativity and other gravity
models can be written in the form \cite{Katana00A}
\begin{equation}                                                  \label{emetok}
  ds^2=|N(q)|(d\tau^2-d\s^2).
\end{equation}
where the conformal factor $N(q)\in {\cal C}^l$, $l\ge2$, is $l$ times
continuously differentiable function of one variable $q\in\MR$ except finite
number of singularities. Let variable $q$ depend only on one coordinate, the
dependence being given by the ordinary differential equation
\begin{equation}                                                  \label{eshiff}
  \left|\frac{dq}{d\z}\right|=\pm N(q),
\end{equation}
with the following sign rule
\begin{equation}                                                  \label{esignr}
\begin{array}{ccl}
N>0: & ~~\z=\s,   &~~\text{$+$ sign (static solution)},\\
N<0: & ~~\z=\tau, &~~\text{$-$ sign (homogeneous solution)}.
\end{array}
\end{equation}
Equations (\ref{emetok}) and (\ref{eshiff}) define four different metrics due to 
the modulus and $\pm$ signs in Eq.(\ref{eshiff}). There are two surfaces with 
static metrics and two surfaces with homogeneous metrics which differ by the 
sign of the derivative $dq/d\z$. Let us denote these domains by the Roman 
numbers,
\begin{equation}                                                  \label{emetdo}
\begin{array}{rcl}
  \text{I}:  ~~& N>0,~~& dq/d\s>0,\\
  \text{II}: ~~& N<0,~~& dq/d\tau<0,\\
  \text{III}:~~& N>0,~~& dq/d\s<0,\\
  \text{IV}: ~~& N<0,~~& dq/d\tau>0.
\end{array}
\end{equation}

To clarify the form of the metric (\ref{emetok}), we note that in domain I
there are Schwarzschild like coordinates $\tau,q$ in which the metric becomes
\begin{equation*}
  ds^2=N(q)d\tau^2-\frac{dq^2}{N(q)}.
\end{equation*}  
So the variable $q$ can be interpreted there as the radius and the conformal
factor as the $g_{00}$ component of the metric.

To transform Lorentzian metric (\ref{emetok}) to the Euclidean signature metric,
we perform the rotation in the complex plane of the coordinate on which the
conformal factor does not depend. The corresponding Riemannian metric is the
solution of the same system of the Euler--Lagrange equations as the original
Lorentzian metric (\ref{emetok}) because the conformal factor does not depend on
this coordinate. This transformation is given by the coordinate changes
$\tau=i\rho$ and $\s=i\rho$ in domains I,III and II,IV, respectively. As a
result, we obtain the metric
\begin{equation}                                                  \label{eemtvi}
  ds^2=-N(q)(d\s^2+d\rho^2),
\end{equation}
where we changed notations $\tau\rightarrow\s$ in domains II and IV. The sign of
the conformal factor $N(q)$ is not fixed, and we consider both positive and
negative definite metrics. After the transformation the variable $q$ depends
only on $\s$, this dependence being given by the ordinary differential equation
\begin{equation}                                                  \label{emdmee}
  \left|\frac{dq}{d\s}\right|=\left|N(q)\right|.
\end{equation}
The modulus signs are opened in the following way
\begin{equation}                                        \label{edomva}
\begin{aligned}
  {\rm I:}   && N>0,~~~~dq/d\s&>0, & \sign g_{\al\bt}&=(--), \\
  {\rm II:}  && N<0,~~~~dq/d\s&<0, & \sign g_{\al\bt}&=(++), \\
  {\rm III:} && N>0,~~~~dq/d\s&<0, & \sign g_{\al\bt}&=(--), \\
  {\rm IV:}  && N<0,~~~~dq/d\s&>0, & \sign g_{\al\bt}&=(++),
\end{aligned}
\end{equation}
where the signature of the metric (\ref{eemtvi}) is shown in the last column.
Metrics in domains I and III as well as in domains II and IV are essentially the
same because they are related by the transformation $\s\rightarrow-\s$.

Riemannian metric defined by Eqs.(\ref{eemtvi}) and (\ref{emdmee}) is the
subject of the present paper. We admit the conformal factor to have zeroes and
singularities at a finite number of points $q_i$, $i=1,\dots,k$. Infinite points
$q_1=-\infty$ and $q_k=\infty$ are included in this sequence. In this way the
real line $q$ is divided on intervals by points $q_i$ in which the conformal
factor is either strictly positive or negative. We consider power behavior of
the conformal factor near the boundary points $q_i$:
\begin{eqnarray}                                        \label{ecfapo}
  |q_i|<\infty:&~~~~~&N(q)\sim|q-q_i|^m,
\\                                                      \label{ecfasi}
  |q_i|=\infty:&~~~~~&N(q)\sim|q|^m.
\end{eqnarray}
For finite $q_i$, the exponent differs from zero, $m\ne0$, because the conformal 
factor either equals zero or singular by assumption. In a general case, the 
exponent $m$ can be an arbitrary real number. Horizons of the space-time 
correspond to zeroes of the conformal factor at finite points $|q_i|<\infty$ in 
the Lorentzian case \cite{Katana90}.

Metric (\ref{eemtvi}) has at least one Killing vector $K=\pl_\rho$, its length
being $-N(q)$.

Christoffel's symbols are defined by the metric,
\begin{equation*}
  \G_{\al\bt}{}^\g
  =g^{\g\dl}(\pl_\al g_{\bt\dl}+\pl_\bt g_{\al\dl}-\pl_\dl g_{\al\bt}),
\end{equation*}
and have the following nontrivial components
\begin{align}                                           \label{ecrson}
  {\rm I,II}:&~~~~~~\G_{\s\s}{}^\s=\G_{\s\rho}{}^\rho=\G_{\rho\s}{}^\rho
  =-\G_{\rho\rho}{}^\s=\frac12N',
\\                                                      \label{ecrstw}
  {\rm III,IV}:&~~~~~~\G_{\s\s}{}^\s=\G_{\s\rho}{}^\rho=\G_{\rho\s}{}^\rho
  =-\G_{\rho\rho}{}^\s=-\frac12N',
\end{align}
where prime denotes the derivative with respect to the argument, $N'=dN/dq$, and
there is no summation over indices $\rho$ and $\s$. The curvature tensor in our
notations is
\begin{equation*}
  R_{\al\bt\g}{}^\dl=\pl_\al\G_{\bt\g}{}^\dl-\G_{\al\g}{}^\e
                     \G_{\bt\e}{}^\dl-(\al\leftrightarrow\bt).
\end{equation*}
It is the same in all domains and has only one independent component
\begin{equation}                                        \label{eriete}
  R_{\s\rho\s}{}^\rho=\frac12NN''
\end{equation}
Nonzero components of the Ricci tensor and the scalar curvature are
\begin{align}                                           \label{eriecl}
  R_{\s\s}&=R_{\rho\rho}=\frac12NN'',
\\                                                      \label{ecusce}
  R&=-N''.
\end{align}
As the consequence, the conformal factor $N$ is the second power polynomial for
constant curvature surfaces $R=\const$.

According to Eq.(\ref{ecusce}), the scalar curvature is singular at $q_i$ for
the following exponents in the power behavior in Eq.(\ref{ecfapo}):
\begin{eqnarray}                                        \label{esfpsc}
  |q_i|<\infty:&~~~~~&m<0,~~0<m<1,~~1<m<2,
\\                                                      \label{esfpcc}
  |q_i|=\infty:&~~~~~&m>2.
\end{eqnarray}
At infinite boundary points $q_i=\pm\infty$ the scalar curvature goes to nonzero
constant for $m=2$ and to zero for $m<0$. Note that the nonzero value of the
scalar curvature at finite points $|q_i|<\infty$ can occur also for $m=1$ due to
the next power corrections in expansion (\ref{ecfapo}).

The range of definition for metric (\ref{eemtvi}) on the $\s,\rho$ plane depends
on the conformal factor. Coordinate $\rho\in\MR$ runs through all real line
because nothing depends on it, but the range of definition of coordinate $\s$
is defined by Eq.(\ref{emdmee}). We have finite, semifinite or infinite interval
for coordinate $\s$ in the each interval $q\in(q_i,q_{i+1})$ depending on
whether the integral
\begin{equation}                                        \label{eferin}
  \s\sim\int^{q_i,q_{i+1}}\frac{dq}{N(q)}
\end{equation}
converge or diverge at the boundary points. Depending on the exponent $m$ we have
\begin{equation}                                        \label{eboucb}
\begin{array}{rl}
  |q_i|<\infty:~~~~&\left\lbrace
  \begin{array}{rll}
  m<1,  ~~&\text{converge,}  \\
  m\ge1,~~&\text{diverge,}
  \end{array}\right. \\
  |q_i|=\infty:~~~~&\left\lbrace
  \begin{array}{rll}
  m\le1,~~&\text{diverge,}\\
  m>1,  ~~&\text{converge,}
  \end{array}\right.
\end{array}
\end{equation}
Coordinate $\s$ runs through all real line, $\s\in(-\infty,\infty)$ if the
integral diverge at both ends of the interval $(q_i,q_{i+1})$, and the metric is
defined on the whole plane $\s,\rho\in\MR^2$. If at one boundary point $q_{i+1}$
or $q_i$ the integral converge, then the metric is defined on the half plane
$\s\in(-\infty,\s_{i+1})$ or $\s\in(\s_i,\infty)$, respectively. The choice of
boundary points $\s_{i+1}$ and $\s_i$ is arbitrary, and without loss of
generality, we can put $\s_{i,i+1}=0$. If the integral converges at both
boundary points, then the solution is defined on the strip
$\s\in(\s_i,\s_{i+1})$, and only one of the ends of the interval can be set to
zero.

To construct the maximally extended surface in the Lorentzian case, we attribute
the conformal block of definite shape to each interval $(q_i,q_{i+1})$ and then
glue them together. For the Euclidean signature metric, there is no need for
this procedure because ``lightlike'' extremals with asymptotics 
$\rho=\pm\s+\const$ are shown to be absent.

The value of variable $q$ and hence the scalar curvature are constant along
Killing trajectories $\s=\const$. Variable $q$ is monotonically increasing on
$\s$ in domains I,IV and monotonically decreasing in domains II,III according to
the definition of the domains (\ref{edomva}).
\section{Extremals                                               \label{sexesm}}
To describe maximally extended surface for metric (\ref{eemtvi}) we must analyze
the behavior of extremals $x^\al(t)=\lbrace\s(t),\rho(t)\rbrace$, $t\in\MR$
given by the system of ordinary differential equations
\begin{equation*}
  \ddot x^\al=-\G_{\bt\g}{}^\al\dot x^\bt\dot x^\g,
\end{equation*}
where dot denotes the derivative with respect to the canonical parameter $t$
which is defined up to a linear transformation. For definiteness, consider
domain I. Expressions for Christoffel's symbols (\ref{ecrson}) yield the system
of equations for extremals
\begin{align}                                        \label{eqexte}
  \ddot\s  &=\frac12N'(\dot\rho^2-\dot\s^2),
\\                                                      \label{eqextt}
  \ddot\rho&= -N'\dot\s\dot\rho.
\end{align}
It has two first integrals
\begin{align}                                           \label{efiice}
  -N(\dot\s^2+\dot\rho^2)&=C_0=\const,
\\                                                      \label{efisce}
  -N\dot\rho&=C_1=\const.
\end{align}
The existence of the integral (\ref{efiice}) allows one to choose the length of
the extremal as a canonical parameter. The second integral (\ref{efisce}) is
connected to the existence of the Killing vector. Now we formulate the theorem
determining extremals.
\begin{theorem}                                         \label{textre}
Any extremal in domain I belongs to one of the four classes.
\newline
1. Straight extremals of the form (the analog of lightlike extremals)
\begin{equation}                                        \label{extlie}
  \rho=\pm\s+\const,
\end{equation}
exist only for the Euclidean metric $N=\const$ and the canonical parameter can
be chosen as $\s=t$.\\
2. General type extremals which form is defined by equation
\begin{equation}                                        \label{extgee}
  \frac{d\rho}{d\s}=\pm\frac1{\sqrt{-1-C_2 N}},
\end{equation}
where $C_2$ is a negative constant. Canonical parameter is defined by any of the
equations
\begin{align}                                           \label{exttpe}
  \dot\s  &=\pm\frac{\sqrt{-1-C_2N}}N,
\\                                                      \label{extgpe}
  \dot\rho&=\frac1N.
\end{align}
The signs plus or minus in Eqs.(\ref{extgee}) and (\ref{exttpe}) must be chosen
simultaneously.\newline
3. Straight extremals parallel to axis $\s$ and going through each point
$\rho=\const$. Canonical parameter is defined by equation
\begin{equation}                                        \label{extste}
  \dot\s=\frac1{\sqrt N}.
\end{equation}
4. Straight degenerate extremals parallel to axis $\rho$ and going through the
points $\s_0=\const$ in which
\begin{equation}                                        \label{extdce}
  N'(\s_0)=0.
\end{equation}
Canonical parameter can be chosen as
\begin{equation}                                        \label{extdge}
  t=\rho.
\end{equation}
\end{theorem}
The proof of this theorem repeats almost word by word the proof of the
corresponding theorem in the Lorentzian case \cite{Katana00A} and is not given
here. Let us remind that constant $C_2$ is defined by the integrals
(\ref{efiice}) and (\ref{efisce})
\begin{equation}                                        \label{econct}
  C_2=\frac{C_0}{C_1^2}.
\end{equation}
Equation (\ref{extgee}) shows that constant $C_2$ parameterizes the angle at
which an extremal of general type goes through a given point.

Behavior of extremals for metrics with Euclidean signature is essentially
different from that for Lorentzian metrics. First, the analog of lightlike
extremals is absent for Riemannian metrics. This is important because lightlike
extremals are incomplete on horizons and must be continued. This problem is
absent in the Riemannian case. Second, Eqs.(\ref{extgee}) and (\ref{exttpe})
differ from the corresponding equations in the Lorentzian case by the sign
before the unity inside the square root. At first glance insignificant, this
difference leads to the absence of ``lightlike'' asymptotics for extremals of
general type as $q\rightarrow q_i$ near zeroes of conformal factor $N(q_i)=0$
which define horizons.

\begin{figure}[p]
\hfill\includegraphics[width=.95\textwidth]{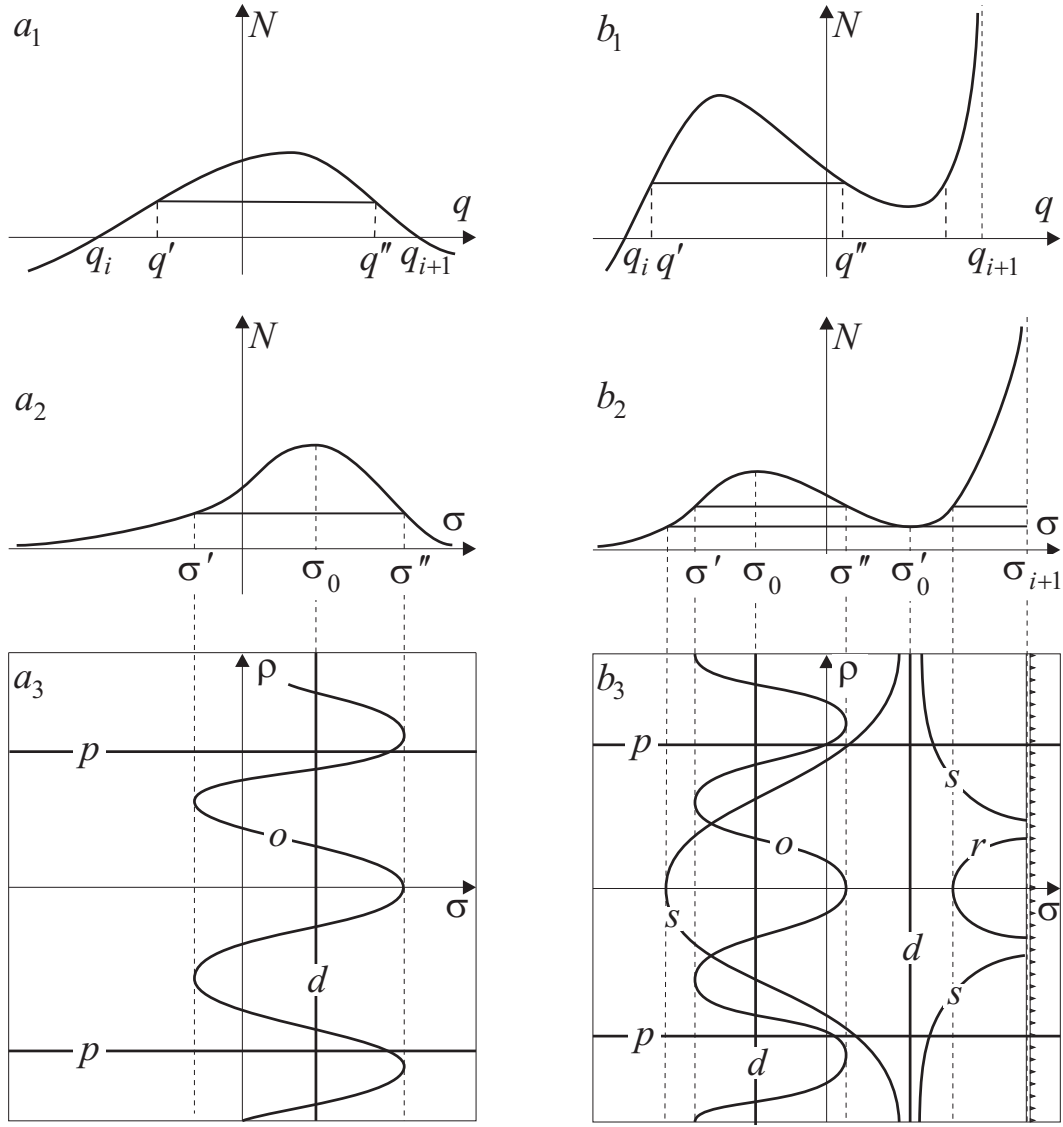}
\hfill {}
\\
\centering \caption{Top row ($a_1$, $b_1$): typical behavior of the conformal
factor $N(q)$ between two zeroes and a zero and singularity. Middle row
($a_2$, $b_2$): dependence of the conformal factor $N(\s)$ on coordinate $\s$.
Point $\s_{i+1}$ can be either at finite or infinite distance from the origin.
Bottom row ($a_3$, $b_3$): extremals of general type ($o$) for different values
of $C_2$ oscillate between $\s'$ and $\s''$ near local maximum. There is one
degenerate extremal ($d$) which goes through every local extremum. There are
extremals ($s$) that asymptotically approach the degenerate extremal at local
minimum $\s'_0$ and can end at the singular boundary $\s_{i+1}$. Extremals of
general type ($r$) of finite length start and end at the singular boundary. All
extremals can be arbitrary  moved along axis $\rho$. Straight extremals
($p$) parallel to axis $\s$ are going through every point $\rho$.
\label{fcoext}}
\end{figure}

Qualitative behavior of extremals of general type is easily analyzed. We
consider domain I for definiteness. The conformal factor is positive in domain
I, and extremals of a general type exist only for negative values of the
constant $C_2$ because otherwise the right hand side of Eq.(\ref{extgpe})
becomes imaginary. The inequality
$$
  N(q)\ge-1/C_2.
$$
must hold. For sufficiently large values of the modulus of $C_2$, this
inequality determines the range of $q\in(q',q'')$ where the boundary points $q'$
and $q''$ are given by equations $N(q)=-1/C_2$. This range definits points $\s'$
and $\s''$ which correspond to $q'$ and $q''$. The extremal of general type can
not go out of the strip $\s\in(\s',\s'')$, $\rho\in(-\infty,\infty)$. Simple
analysis of Eq.(\ref{extgpe}) shows that extremals of general type oscillate
between the values $\s'$ and $\s''$ as shown in Fig.\ref{fcoext}$a_3$,$b_3$.
Oscillating extremals of general type are always complete because the right hand
side of Eq.(\ref{extgpe}) is bounded from the top by $1/N(q')$ or $1/N(q'')$ and
bottom by $1/\max N(q)$.

If the conformal factor is equal to infinity in point $q_{i+1}$ as shown in
Fig.\ref{fcoext}$b$, then an extremal of general type can start and end at the
singular boundary. This boundary corresponds to finite value $\s_{i+1}$ for
$|q_{i+1}|<\infty$, $m<1$ and $|q_{i+1}|=\infty$, $m>1$. All these extremals
approach the singular boundary at right angle because the right hand side of
Eq.(\ref{extgee}) goes to zero. Completeness of extremals of general type going
to the singular boundary is defined by the integral
\begin{equation}                                                  \label{extgpt}
  \lim_{q\to q_{i+1}}t\to\int^{q_{i+1}}\frac{dq}{\sqrt{N}},
\end{equation}
as the consequence of Eqs.(\ref{exttpe}) and (\ref{emdmee}). So completeness
of extremals of general type which approach the singular boundary is the same as
completeness of straight extremals parallel to axis $\s$ (\ref{extste}). They
are complete only for $|q_{i+1}|=\infty$ and $1<m\le2$.

Typical behavior of the conformal factor between two zeroes and with two local
extrema between zero and singularity is shown in the top row of
Fig.\ref{fcoext}, ($a_1$, $b_1$). In case 1$b$ we assume $|q_{i+1}|<\infty$ and 
$m<1$ or $1<m<2$ corresponding to the curvature singularity. In the middle row 
($a_2$, $b_2$), we shaw the dependence of the conformal factors on coordinate
$\s$. The value of coordinate $\s_{i+1}$ is finite for $m<1$ and infinite for
$1<m<2$ in Fig.\ref{fcoext},$b_2$. Qualitative behavior of extremals on the
$\s,\rho$ plane is shown in the bottom row ($a_3$, $b_3$). Extremals of general
type $(o)$ oscillate near local maximum $N(\s_0)$ between $\s'$ and $\s''$ which
are determined by the value of constant $C_2$. These extremals $(r)$ can also
start and end at the singular boundary $\s_{i+1}$ having a finite length.
Degenerate extremals $(d)$ go through every extremum of the conformal factor.
There are also extremals of general type $(s)$ which asymptotically approach the
degenerate extremal going through local minimum $\s'_0$ when
$\rho\to\pm\infty$. Part of these extremals end at the singular boundary
$\s_{i+1}$ at a finite value of the canonical parameter. All extremals can be
shifted arbitrary along axis $\rho$. Straight extremals $(p)$ parallel to axis
$\s$ go through every point $\rho$.

If both boundary points $q_i$ and $q_{i+1}$ of the interval are zeroes, then
the conformal factor $N$ has at least one maximum as the consequence of
continuity through which goes the degenerate extremal ($d$). Degenerate
extremals are always complete, i.e.\ have infinite length, because the canonical
parameter coincides with coordinate $\rho$ (\ref{extdge}).

The above analysis shows that incompleteness of extremals in strips
$\s\in(\s_i,\s_{i+1})$ and $\rho\in(-\infty,\infty)$ is defined entirely by
completeness of straight extremals parallel to axis $\s$ when they approach
boundary points $\s_i,\s_{i+1}$. In its turn, this is defined by the convergence
of the integral (\ref{extgpt}). These extremals are incomplete at finite points
$|q_i|<\infty$ for $m<2$. At infinite points $|q_i|=\infty$ they are incomplete
for $m>2$ and complete in all other cases. Extension of the surface is necessary
only for the conformal factor which has simple zero $m=1$ at a finite point
$q_i$ because we must extend the surface only for nonsingular curvature. Note
that a simple zero in the Lorentzian case corresponds to a horizon, and this is
the unique case when four conformal blocks meet at a saddle point.
The Carter--Penrose diagram for the Schwarzschild solution has exactly this form.
\section{Global solutions}
To construct maximally extended surfaces with metric (\ref{eemtvi}), we perform
the following procedure which is useful also for visualization of the
surfaces. We identify points $\rho$ and $\rho+L$ where $L$ is an arbitrary
positive constant which is always possible because nothing depends on coordinate
$\rho$. After this identification the plane $\s,\rho$ becomes a cylinder. The
circumference of the directing circle is equal to
\begin{equation}                                                  \label{elenci}
  P=LN(q)\rightarrow
\begin{cases} 0, & N(q_i)=0, \\ \infty & N(q_i)=\infty. \end{cases}
\end{equation}
The plane $\s,\rho$ is the universal covering space for this cylinder.

We summarize properties of the boundary points $q_i$ in Table \ref{tprobo}.
\begin{table}
\begin{center}
\begin{tabular}{|c|c|c|c|c|c|c|}
     \multicolumn{7}{c}{$|q_i|<\infty$}\\ \hline
     & $m<0$ & $0<m<1$ & $m=1$ & $1<m<2$ & $m=2$ & $m>2$ \\ \hline
$R$    & $\infty$ & $\infty$ & $\const$ & $\infty$ & $\const$ & $0$ \\
$\s_i$ & $\const$ & $\const$ & $\infty$ & $\infty$ & $\infty$ & $\infty$ \\
$P$    & $\infty$ & $0$ & $0$ & $0$ & $0$ & $0$ \\
Completeness& $-$ & $-$ & $-$ & $-$ & $+$ & $+$ \\  \hline
     \multicolumn{7}{c}{$|q_i|=\infty$}\\ \hline
     & $m<0$ & $m=0$ & $0<m\le 1$ & $1<m<2$ & $m=2$ & $m>2$ \\ \hline
$R$    & $0$ & $0$ & $0$ & $0$ & $\const$ & $\infty$ \\
$\s_i$ & $\infty$ & $\infty$ & $\infty$ & $\const$ & $\const$ & $\const$ \\
$P$    & $0$ & $\const$ & $\infty$ & $\infty$ & $\infty$ & $\infty$ \\
completeness& $+$ & $+$ & $+$ & $+$ & $+$ & $-$ \\ \hline
\end{tabular}
\end{center}
 \caption{Properties of boundary points depending on the exponent $m$. The
 symbol $\const$ in the rows for the scalar curvature denotes a nonzero constant.
 \label{tprobo}}
\end{table}
Depending on the value of the exponent $m$, we show there the values of scalar 
curvature $R$ on the corresponding surface, finiteness of coordinate $\s_i$ at 
the point $q_i$, circumference of directing circles for cylinders $P$ and 
completeness of extremals which are parallel to axis$\s$.

Forms of the surfaces near boundary points $q_{i+1}$ after the identification
$\rho\sim\rho+L$ are shown in Fig.\ref{fsurec}. Surfaces near points $q_i$ have
similar form but are turned in the opposite direction. The surface
corresponding to the whole interval $(q_i,q_{i+1})$ is obtained by gluing two
such surfaces for boundary points $q_i$ and $q_{i+1}$ together.
\begin{figure}[p]
\hfill\includegraphics[width=.95\textwidth]{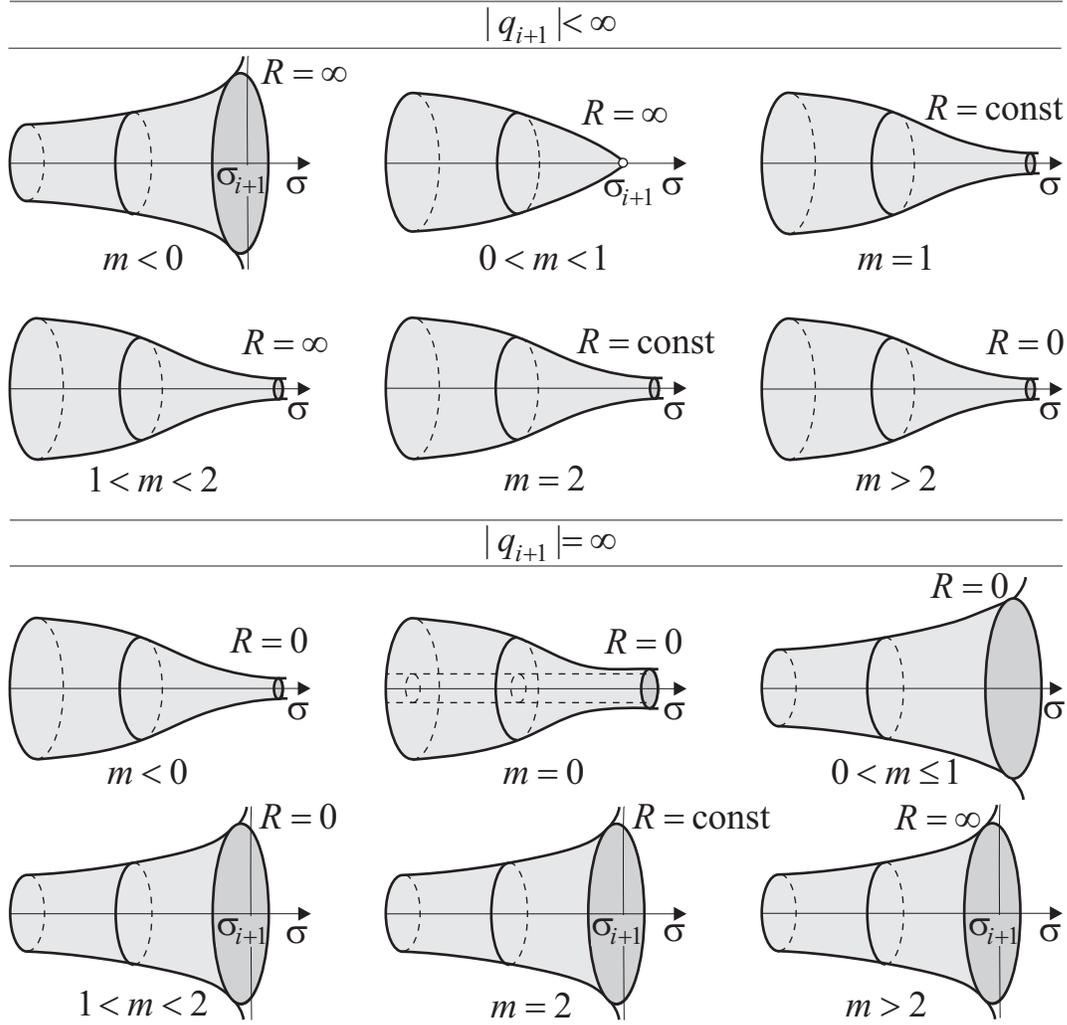}
\hfill {}
\\
\centering
 \caption{Forms of the surfaces near boundary points $q_{i+1}$ after the
 identification $\rho\sim\rho+L$. By assumption, the coordinate $\s$ grows from
 left to right and $\s_{i+1}>\s_i$. Surfaces near points $q_i$ have a similar
 form but are turned in the opposite direction. The surface for the whole
 interval $q_i,q_{i+1}$ is obtained by gluing two such surfaces for boundary
 points $q_i$ и $q_{i+1}$ together.
 \label{fsurec}}
\end{figure}
So extremals must be continued only near the boundary point $|q_i|<\infty$ for
$m=1$. We call this point horizon because it corresponds to a horizon in the
Lorentzian case. The continuation at a horizon is performed as follows. First of
all note that a horizon in the Euclidean case is itself a point because the
length of the directing circle goes to zero. Next, this ``infinite'' point in
the $\rho,\s$ plane lays, in fact, at a finite distance because all extremals
reach this point at a finite value of the canonical parameter. Up to higher
order terms, the conformal factor in the neighborhood of the point $q_i$ has
the form
\begin{equation}                                        \label{eclcod}
  N(q)=N'_i(q-q_i),
\end{equation}
where $N'_i=N'(q_i)=\const\ne0$. In domain I for $N'_i>0$ and $q>q_i$,
Eq.(\ref{emdmee}) is easily integrated
$$
  q-q_i=e^{N'_i\s},
$$
where we dropped an insignificant constant of integration related to a shift of
$\s$. Thus the boundary point $|q_i|<\infty$ is reached for
$\s\rightarrow-\infty$. Metric (\ref{eemtvi}) in coordinates $q,\rho$ which play
the role of Schwarzschild coordinates takes the form
\begin{equation}                                        \label{eclmes}
  -ds^2=\frac{dq^2}{N'_i(q-q_i)}+N'_i(q-q_i)d\rho^2.
\end{equation}
In polar coordinates $r,\vf$, defined by the transformation
\begin{equation}                                        \label{epocoe}
  q-q_i=\frac{N'_i}4r^2,~~~~~~\rho=\frac2{N'_i}\vf,
\end{equation}
the metric becomes Euclidean
$$
  -ds^2=dr^2+r^2d\vf^2.
$$
Here the polar angle varies within the interval $\vf\in(0,LN'_i/2)$ and the
radius $r$ is defined in the neighborhood of the boundary point $q_i$ by
Eq.(\ref{epocoe}). This coordinate transformation maps the ``infinite'' line
$\s_i=-\infty$, $\rho\in\MR$ into the origin of Euclidean plane. Conical 
singularity can appear at the origin because the polar angle varies within the 
interval which in general differs from $(0,2\pi)$. The corresponding deficit 
angle is
\begin{equation}                                        \label{edefae}
  2\pi\theta=\frac{LN'_i}2-2\pi.
\end{equation}
Thus we get the Euclidean metric on a plane $\MR^2$ with a conical singularity
at the origin. The deficit angle is zero for $L=4\pi/N'_i$, conical singularity
is absent, and we are left with the flat Euclidean metric which is evidently
smooth at the origin. In general, conformal factor (\ref{eclcod}) has
corrections of higher order near the boundary point $q_i$ and transformation to
polar coordinates yields the metric of the same differentiability as the
conformal factor.

In general continuation of the solution through the point $|q_i|<\infty$ for
$m=1$ has no meaning because this point correspond to a conical singularity.
We assume that this point as well as any other singular point does not belong
to a manifold. Therefore the plane $\s,\rho$ or its part is the universal
covering space for the surface with metric (\ref{eemtvi}). Continuation is
necessary only in the absence of conical singularity $L=4\pi/N'_i$. In this case,
straight extremals parallel to axis $\s$ and going through points $\rho$ and
$\rho+L/2$ become two halves of the same extremal shown in Fig.\ref{fcorec}.
The fundamental group is trivial in the absence of a conical singularity, and
therefore the corresponding surface is itself a universal covering space.
\begin{figure}[htb]
\hfill\includegraphics[width=.35\textwidth]{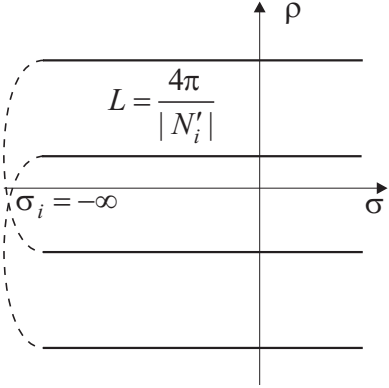}
\hfill {}
\\
\centering
 \caption{Continuation of straight extremals going through points $\rho$ and
 $\rho+L/2$ in the absence of conical singularity $L=4\pi/|N'_i|$. The
 identification is performed in the point $\s_i=-\infty$.}
 \label{fcorec}
\end{figure}

\begin{figure}[htb]
\hfill\includegraphics[width=.95\textwidth]{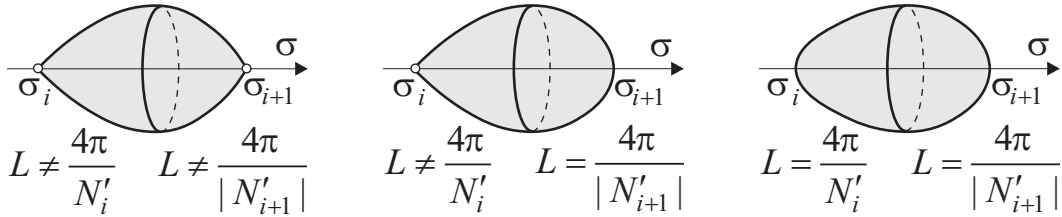}
\hfill {}
\\
\centering
 \caption{Three types of possible maximally extended surfaces, corresponding to
 the interval $(q_i,q_{i+1})$ when the conformal factor has asymptotics
 $N\sim(q-q_{i,i+1})$ in boundary points. In general, the surface has two
 conical singularities after the identification $\rho\sim\rho+L$. For
 $L|N'_{i,i+1}|=4\pi$ conical singularities are absent.
 \label{fglmon}}
\end{figure}

If the conformal factor has asymptotics $N\sim(q-q_i)$ and $N\sim(q_{i+1}-q)$ on
both sides of the interval $(q_i,q_{i+1})$, then the surface must be continued
in both points $\s=\pm\infty$. In general, after the identification
$\rho\sim\rho+L$ the surface has conical singularities in both points. For
$LN'_i=4\pi$ and $LN'_{i+1}=4\pi$ conical singularities are absent. There are
three types of global surfaces shown in Fig.\ref{fglmon} according to the number
of conical singularities. These surfaces have topology of a cylinder, plane or
sphere, respectively.

The rules for construction of maximally extended surfaces with metric
(\ref{eemtvi}) are as follows.
\begin{enumerate}
\item The maximally extended surface corresponds to each interval
$(q_i,q_{i+1})$ after the identification $\rho\sim\rho+L$ and is obtained by
gluing of two surfaces shown in Fig.\ref{fsurec} corresponding to the boundary
points $q_i$ and $q_{i+1}$.
\item In all cases except the absence of the conical singularity,
$|q_i|<\infty$, $LN'_i\ne4\pi$ or $|q_{i+1}|<\infty$, $L|N'_{i+1}|\ne4\pi$ the
strip $\s\in(\s_i,\s_{i+1})$, $\rho\in\MR$ with metric (\ref{eemtvi}) is the
universal covering space for the corresponding maximally extended surface.
\item In the absence of one of conical singularities, $|q_i|<\infty$,
$LN'_i=4\pi$ or $|q_{i+1}|<\infty$, $L|N'_{i+1}|=4\pi$, the surface obtained
from the plane $\s,\rho$ by identification $\rho\sim\rho+L$ is itself
the maximally extended surface with trivial fundamental group.
\end{enumerate}

Note that we do not glue together coordinate charts. It means that the
corresponding surfaces are smooth $\CC^\infty$. In the absence of conical
singularity, the transformation to polar coordinates (\ref{epocoe}) does not use
the explicit form of the conformal factor, and the resulting surface is also
smooth as the Euclidean plane. So we have proved the statement.
\begin{theorem}                                                   \label{tunisp}
The universal covering space constructing according to the rules 1--3 is the
maximally extended smooth surface, $\CC^\infty$, with Riemannian
${\cal C}^l$, $l\ge2$, metric such that every point not lying on a horizon has a
neighborhood isometric to some domain with metric (\ref{eemtvi}).
\end{theorem}

The universal covering space is known to be unique and all other maximally
extended surfaces are obtained as quotient spaces of the universal covering
space by the transformation group which acts freely and properly discontinuous
\cite{KobNom6369}.
\section{Schwarzschild solution}
\begin{figure}[h,b,t]
\hfill\includegraphics[width=.95\textwidth]{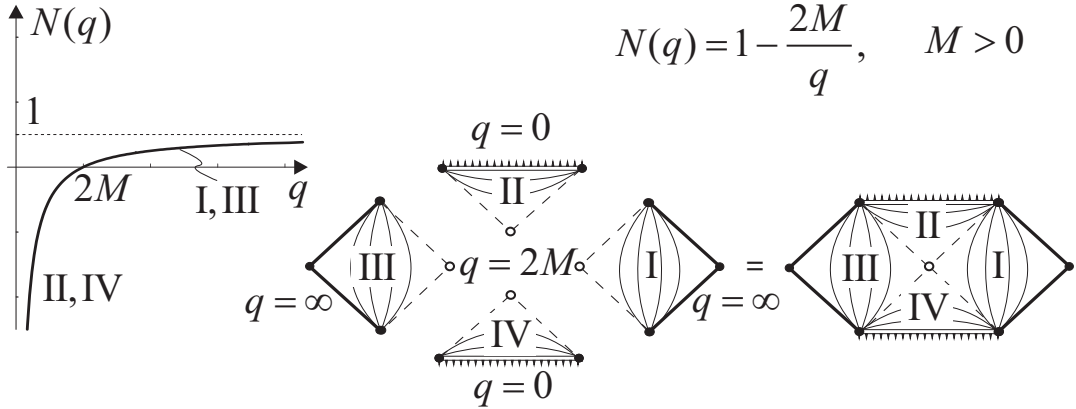}
\hfill {}
\\
\centering \caption{Construction of the Carter--Penrose diagram for the
Schwarzschild solution by the method of conformal blocks.\label{fsheso}}
\end{figure}

We consider the Schwarzschild solution as an example. Look for spherically
symmetric solutions of vacuum Einstein's equations in the form
\begin{equation*}
  ds^2=g_{\mu\nu}dx^\mu dx^\nu=f(d\tau^2-d\s^2)-m(d\theta^2+\sin^2\theta d\vf^2),
\end{equation*}
where $f(\tau,\s)$ and $m(\tau,\s)$ are two unknown functions on time $\tau$ and
radius $\s$, $\mu,\nu=0,1,2,3$. Then, as the consequence of the equations of
motion, a solution depends only on one of the coordinates (Birkhoff's theorem)
and has the form
\begin{equation}                                                  \label{eschco}
  ds^2=|N(q)|(d\tau^2-d\s^2)-q^2(d\theta^2+\sin^2\theta d\vf^2),~~~~q>0,
\end{equation}
where
\begin{equation*}
  N(q)=1-\frac{2M}q,~~~~q=\begin{cases} r, & N(q)>0 \\ t, & N(q)<0.
\end{cases}
\end{equation*}
Detailed calculations are given in \cite{KaKlKu99}. The variable $q$ is related
to one of the coordinates $\tau$ or $\s$ by differential Eq.(\ref{eshiff}). So
the time-radial part of the Schwarzschild metric coincides exactly with metric
(\ref{emetok}), (\ref{eshiff}). The appearance of the modulus and $\pm$ signs
is the consequence of Einstein's equations. Maximally extended Schwarzschild
solution is well known \cite{Kruska60,Szeker60} and was obtained by introducing
global coordinates. In general, introduction of such coordinates is not
necessary and can be very complicated. The method of conformal blocks for
construction of global solutions for two-dimensional metrics (\ref{eshiff}) was
proposed in \cite{Katana00A}. Topologically, the maximally extended
Schwarzschild solution is the direct product of two surfaces $\MU\times\MS^2$
where $\MS^2$ is a sphere and $\MU$ is the two-dimensional Lorentzian surface
which is represented by the Carter--Penrose diagram shown in Fig.\ref{fsheso}.
The dependence of the conformal factor $N$ on $q$ is shown in the figure. It has
one simple zero at $q=2M$. Thus the interval $(0,\infty)$ is divided on two
intervals $(0,2M)$ and $(2M,\infty)$ where the conformal factor is negative
(homogeneous solutions) and positive (static solutions). Two triangular
conformal blocks II, IV  and two square conformal blocks I, III correspond to
intervals $(0,2M)$ and $(2M,\infty)$, respectively. Their unique gluing yields
the Carter--Penrose diagram for the Schwarzschild solution and is shown in
Fig.\ref{fsheso}. The rules for construction of conformal blocks, their gluing
procedure, and the proof of the differentiability of the metric on the glued
boundaries are given in \cite{Katana00A}.
\begin{figure}[h,b,t]
\hfill\includegraphics[width=.95\textwidth]{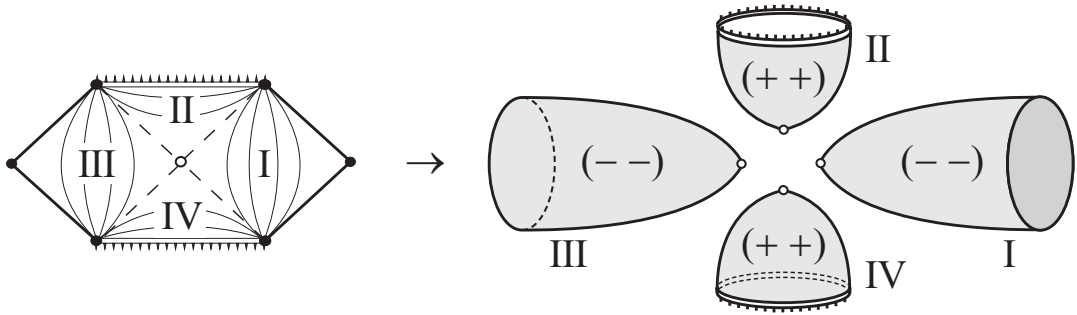}
\hfill {}
\\
\centering \caption{The Carter--Penrose diagram for the Schwarzschild solution
in the Euclidean case breaks into four disconnected between themselves surfaces.
There are two surfaces I, III with negative definite metric and two surfaces II,
IV with positive definite metric corresponding to the regions outside and inside
the horizon, respectively. \label{feushw}}
\end{figure}

The Carter--Penrose diagram after changing the signature from Lorentzian to
Euclidean breaks into four disconnected between themselves surfaces which are
shown in Fig.\ref{feushw}. Two Riemannian surfaces I and III with negative
definite metric correspond to the region outside the black hole. For these
surfaces the total metric has the signature $\sign g_{\mu\nu}=(----)$. Negative
definiteness of the metric is due to the choice of the signature for the
Schwarzschild solution (\ref{eschco}) and can be easily changed. This solution
is usually considered as the Euclidean version of the Schwarzschild solution.
Two surfaces II and IV with positive definite metric correspond to regions
inside the horizon of the black hole. The signature of the total metric is
$\sign g_{\mu\nu}=(++--)$. This spherically symmetric solution is usually
considered as unphysical though there is no mathematical reason to discard it.
\section{Conclusion}
We formulated the method of construction of global (maximally extended) surfaces
for a large class of two-dimensional Riemannian metrics having one Killing
vector. The smoothness of surfaces and differentiability of metrics are proved.
This method is the counterpart of the method of conformal blocks for metrics
with Lorentzian signature. Almost anytime each conformal block is the universal
covering space for the maximally extended surface. The exceptions are surfaces
with a horizon which is a point in the Euclidean case. If there is no conical
singularity at this point then the universal covering space appears after the
necessary identification of some points of the conformal block.

The transformation from Euclidean signature of the metric to Lorentzian one is
interesting. If the Lorentzian surface is represented by the Carter--Penrose
diagram with horizons corresponding to zeroes of the conformal factor, then,
after the rotation to the Euclidean signature, the Carter--Penrose diagram
breaks into disconnected surfaces with positive and negative definite metrics.
The change in the signature of the Riemannian metric corresponds to crossing the
horizon in the Lorentzian case.

Two-dimensional metrics considered in the present paper arise not only in
two-dimensional models but also in higher dimensional gravity when solutions
depend essentially only on two of the coordinates. The method can be useful in
this case as well.

The author is very grateful to Prof.~W.~Kummer for numerous and fruitful
discussions during which the idea of this work was born in the middle of
90-ties.

\end{document}